\documentclass{iau}

\usepackage{amsmath}
\usepackage{graphicx}
\usepackage{multirow}

\def\araa{ARAA}
\def\apj{ApJ}

\def\apjs{ApJS}

\def\aap{A\&A}

\def\mnras{MNRAS}

\def\physrep{Phys.~Rep.}

\begin{document}

\lefttitle{Simon Selg}
\righttitle{Proceedings of the International Astronomical Union: Interacting Galaxies}

\jnlPage{1}{7}
\jnlDoiYr{2022}
\doival{10.1017/xxxxx}

\aopheadtitle{Proceedings IAU Symposium}
\editors{The editors}

\title{Proceedings of the International Astronomical Union: Studying Magnetic Field Amplification in Interacting Galaxies Using Numerical Simulations}

\author{Simon Selg, Wolfram Schmidt}
\affiliation{Hamburger Sternwarte, Universit\"{a}t Hamburg, Gojenbergsweg 112, D-21029 Hamburg, Germany \email{simon.selg@hs.uni-hamburg.de}}

\begin{abstract}
There are indications that the magnetic field evolution in galaxies might be massively shaped by tidal interactions and mergers between galaxies. The details of the connection between the evolution of magnetic fields and that of their host galaxies is still a field of research. 


We use a combined approach of magnetohydrodynamics for the baryons and an N-body scheme for the dark matter to investigate magnetic field amplification and evolution in interacting galaxies.

We find that, for two colliding equal-mass galaxies and for varying initial relative spatial orientations, magnetic fields are amplified during interactions, yet cannot be sustained. Furthermore, we find clues for an active mean-field dynamo.
\end{abstract}

\begin{keywords}
galaxies: interactions, turbulence, magnetic fields, MHD, methods: numerical
\end{keywords}

\maketitle

\section{Introduction}

It remains unclear how primordial magnetic fields of order $\ll 10^{-9}\;$G are amplified inside galaxies to values of order more than $10^{-6}\;$G around $z=0$. Possible solutions are the Biermann battery, a dynamo or galaxy interactions \citep[e.g.][and references therein]{Beck+1996, Beck2015, Brandenburg+2005, Subramanian2016}. The generation and action of dynamos has been studied in simulations of isolated galaxies \citep[e.g.][]{Ntormousi+2020, Schober+2013, Steinwandel+2019}.

Interactions between and mergers of galaxies are assumed to be an integral part of galaxy evolution. Within the framework of $\Lambda$CDM it is assumed that hierarchical structure formation leads to larger galaxies being formed through subsequent mergers of smaller galaxies \citep[e.g.][and references therein]{Helmi2020}.

\cite{Drzazga+2011} give evidence for temporally enhanced magnetic fields in observational data. They find that the magnetic field strength in interacting galaxies -- both at close encounters and especially at coalescence -- is enhanced by a factor of 2-3. Simulations performed by \citet{Kotarba+2010} with the SPH code Gadget have confirmed this.  \citet{Renaud+2015, Renaud+2019} investigate starbursts in hydrodynamical simulations using adaptive mesh refinement (AMR) and \citet{Rodenbeck+2016} simulate the evolution of magnetic fields in interacting galaxies using the AMR N-body code Enzo \citep{Bryan+2014}.

However, \citet{Rodenbeck+2016} lack a correct treatment of the dark matter. We therefore model magnetic field amplification in interacting galaxies subject to varying initial conditions, with each galaxy residing in a live dark matter halo.

In Section~\ref{sec:num} we introduce our numerical model and the suite of simulations we performed while in Section~\ref{sec:results} we highlight and discuss some of the results. 

\section{Numerical Methods}\label{sec:num}

\subsection{Galaxies}
Galaxy mergers can be simulated either in large quantities within cosmological simulations \citep[e.g.][]{Patton+2020} or much smaller quantities, i.e. mostly pair-wise \citep[e.g.][but see \cite{Kotarba+2011} for the exception of three galaxies]{Rodenbeck+2016}. For reasons of higher numerical resolution, we choose to use a $2^3\;\mathrm{kpc}^{3}$ computational domain with one pair of galaxies. 

Different from using galaxy formation recipes of cosmological galaxy formation \citep[e.g. ][]{Pillepich+2018} we choose an approach to initialise galaxies a priori. This gives us more control over the exact choice of initial conditions for both galaxies. In our model each galaxy is comprised of a gas disk which is embedded inside a dark matter halo. We use a modified version of the methods developed by \citet{Rodenbeck+2016} and \citet{Wang+2010} in order to initialise a three-dimensional gas disk in equilibrium that is magnetised and hosted by a dark matter halo. While \citet{Wang+2010} have introduced an equilibrium disk model that supports the addition of external potentials to represent the gravitational influence of stars and dark matter it has been extended by \citet{Rodenbeck+2016} to include magnetic fields. Since \citet{Rodenbeck+2016} do not have dark matter halos in their simulations of galaxy interactions we extend their model and that of \citet{Wang+2010} to include live dark matter halos which are sampled via an N-body simulation of $2\times 10^{6}$ particles per halo. 

\citet{Wang+2010} describe how to compute the density profile and the rotation curve using an iterative approach. \citet{Rodenbeck+2016} include magnetic fields:
\begin{equation}
    B = B_{\mathrm{tor}} = \sqrt{8\pi\epsilon_{\mathrm{mag}}\rho\;c_{\mathrm{s}}^{2}},
\end{equation}
where $\epsilon_{\mathrm{mag}}$ is the inverse of the plasma beta which describes the ratio of thermal and magnetic pressure. 

\citet{Wang+2010} show that the velocity profile of the entire three-dimensional disk can be specified in the midplane \begin{equation}
    v_{\mathrm{rot}}^{2}(r,z) = v_{\mathrm{thin}}^{2}(r) + (1+\epsilon_{\mathrm{mag}})c_{\mathrm{s}}^{2}\frac{\partial\ln{\rho}}{\partial\ln{r}}\bigg\vert_{\mathrm{z=0}},\label{eq:rotation01}
\end{equation}
where the first term on the right hand side is described by a thin disk approximation \citep[see for details e.g.][]{BinneyTremaine2008, Freeman1970} including the additional magnetic pressure introduced by \citet{Rodenbeck+2016}. The gas density distribution is computed with an iterative scheme proposed by \citet{Wang+2010}: 
\begin{equation}
    \rho_{\mathrm{gas}}(r, z) = \rho_{0}(r)\exp{\left(-\frac{\Phi_{\mathrm{z}}(r,z)}{(1+\epsilon_{\mathrm{mag}})c_{\mathrm{s}}^{2}}\right)}.
\end{equation}
Here, $\Phi_{\mathrm{z}}$ denotes the vertical difference of the total gravitational potential ($\Phi = \Phi_{\mathrm{gas}} + \Phi_{\mathrm{DM}}$) with respect to its midplane value. 

The dark matter halo is modelled by the following density distribution \citep{Hernquist1990}: 
\begin{equation}
    \rho_{\mathrm{DM}} = \frac{M_{\mathrm{DM}}}{2\pi}\frac{a}{r}\frac{1}{(r+a)^{3}},
\end{equation}
with the scale length $a$. Furthermore, a connection between the Hernquist profile and an NFW profile \citep{Navarro+1996} is established by relating the scale lengths $a$ and $r_{\mathrm{s}}$ as \cite{Springel+2005} propose:
\begin{equation}
    a = r_{\mathrm{s}}\sqrt{2(\ln{(1+c)} - c/(1+c))},\label{eq:halo_04}
\end{equation}
where $c$ is the concentration parameter. We use the method developed by \citet{Drakos+2017} to find a stable N-body representation of a live dark matter halo. 

\subsection{Simulations}
Inside the cosmic web galaxies move on orbits of different shape and orientation. We try to account for this fact in our simulations by performing a parametric study where we vary the relative orientations between the galaxies by an inclination angle $i$ and an impact parameter which can also be described by an angle $\alpha_{\mathrm{b}}$. The relative inclination is computed from the individual galaxies' inclinations: $i = |i_{2}-i_{1}|$. Each galaxy's inclination is computed as the angle between its angular momentum vector and the relative bulk velocity vector, $\mathbf{v} = \mathbf{v}_{\mathrm{rel}} = \mathbf{v}_{1} - \mathbf{v}_{2}$:
\begin{equation}
    i_{\mathrm{j}} = \angle{(\mathbf{L}_{\mathrm{j}}, \mathbf{v}_{\mathrm{rel}})},\ j\in\{1,2\}.
\end{equation}
Additionally, spatial orientation is determined through an impact parameter $b$ which is the vertical distance between the centre of the primary galaxy and the line in the direction of the initial velocity of the secondary galaxy. At the beginning of each simulation, the two galaxies have a separation corresponding to the cut-off radius ($d_{\mathrm{sep}} = 209\;$kpc) of a dark matter halo. Thus, $\sin{\alpha_{\mathrm{b}}} = b/d_{\mathrm{sep}}$ defines the angle associated with the impact parameter. In our simulations we realised a total of 26 different initial configurations covering inclinations between $0^{\circ}$ and $90^{\circ}$ and impact parameter angles between $0^{\circ}$ and $45^{\circ}$.
Each galaxy consists of a gas disk of $10^{10}\;\mathrm{M}_{\odot}$ and a dark matter halo of $10^{12}\;\mathrm{M}_{\odot}$. The radial scale length is $3.5\;$kpc following the galaxy description in \citet{Wang+2010}. We solve the equations of ideal MHD and the evolution of the live halos with the AMR \& N-body code Enzo \citep{Bryan+2014} and employ a subgrid-scale model for unresolved MHD turbulence \citep{Grete+2017, Grete+2019}. An adiabatic equation of state is used. 

\section{Results \& Discussion}\label{sec:results}
\begin{figure}[tb]
  \centerline{\vbox to 6pc{\hbox to 10pc{}}}
  \includegraphics[scale=.4]{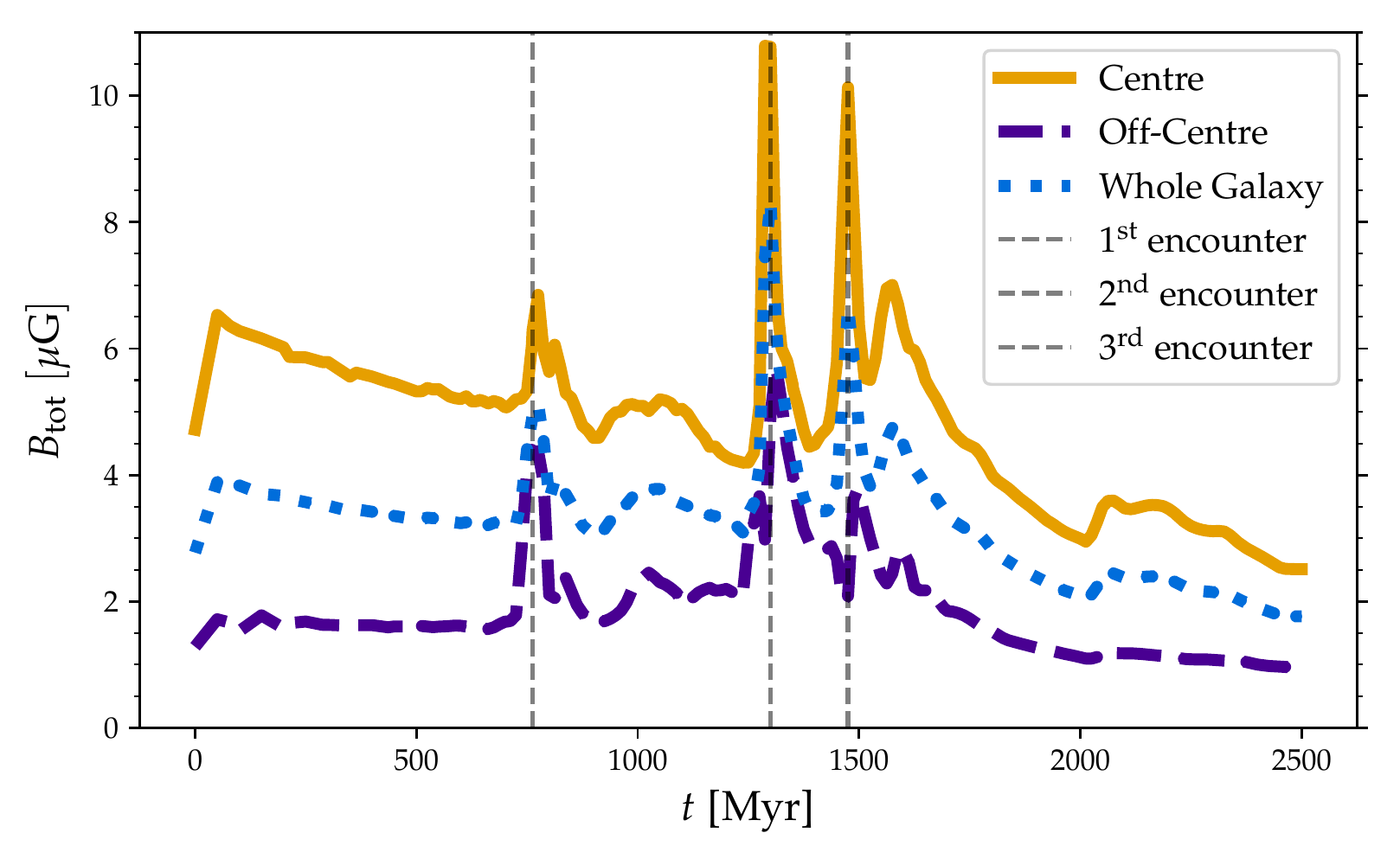}
  \includegraphics[scale=.4]{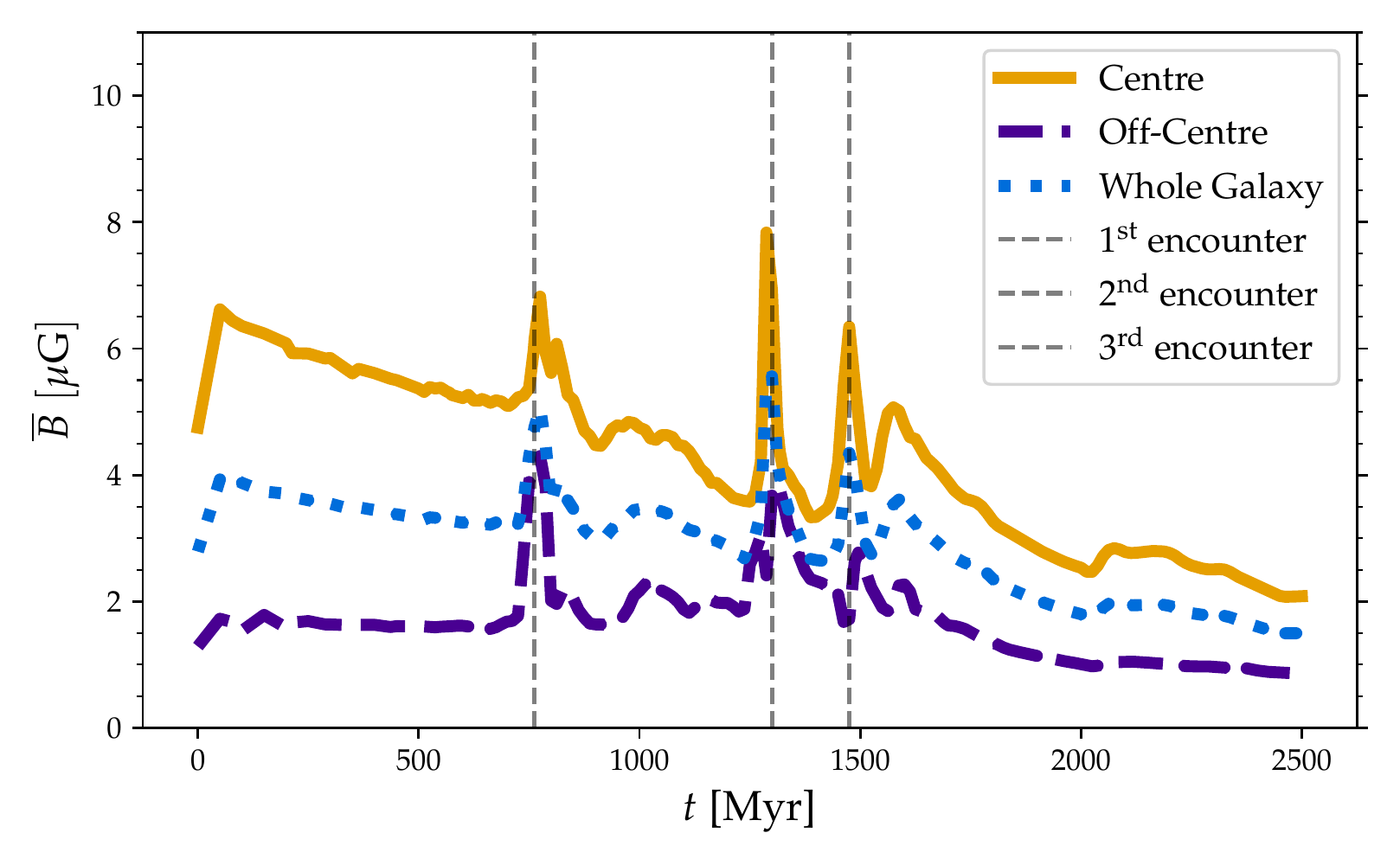}
  \caption{Timeseries of magnetic field evolution in three regions of the primary galaxy: the total magnetic field strength $B_{\mathrm{tot}}$ and the filter-averaged magnetic field strength $\overline{B}$ (see Eq.~\ref{eq:mfe}) are displayed on the left and right image, respectively. We highlight times of three encounters between the two galaxies at $762.5$, $1300$, and $1475\;$Myr.}
  \label{fig:MagneticFieldStrength}
\end{figure}
In the following, we limit our presentation of results to those of one simulation in particular: $i_{1}=i_{2}=90^{\circ}$ and $\alpha_{\mathrm{b}}=20^{\circ}$\footnote{We plan the presentation of results of the other simulations in a future publication.}. For our analysis, trajectories of both galaxies are computed and at the centre of mass of the primary galaxy a cylindrical region of $20\;$kpc radius and $8\;$kpc thickness is positioned. In a fashion similar to \citet{Drzazga+2011} we further subdivide this region into a central cylinder of $5\;$kpc radius -- the centre region -- and a thick shell, $5<r\leq 20\;$kpc -- the off-centre region. Within these regions mass-weighted quantities are computed.
\begin{figure}[tb]
  \centerline{\vbox to 6pc{\hbox to 10pc{}}}
  \includegraphics[scale=.4]{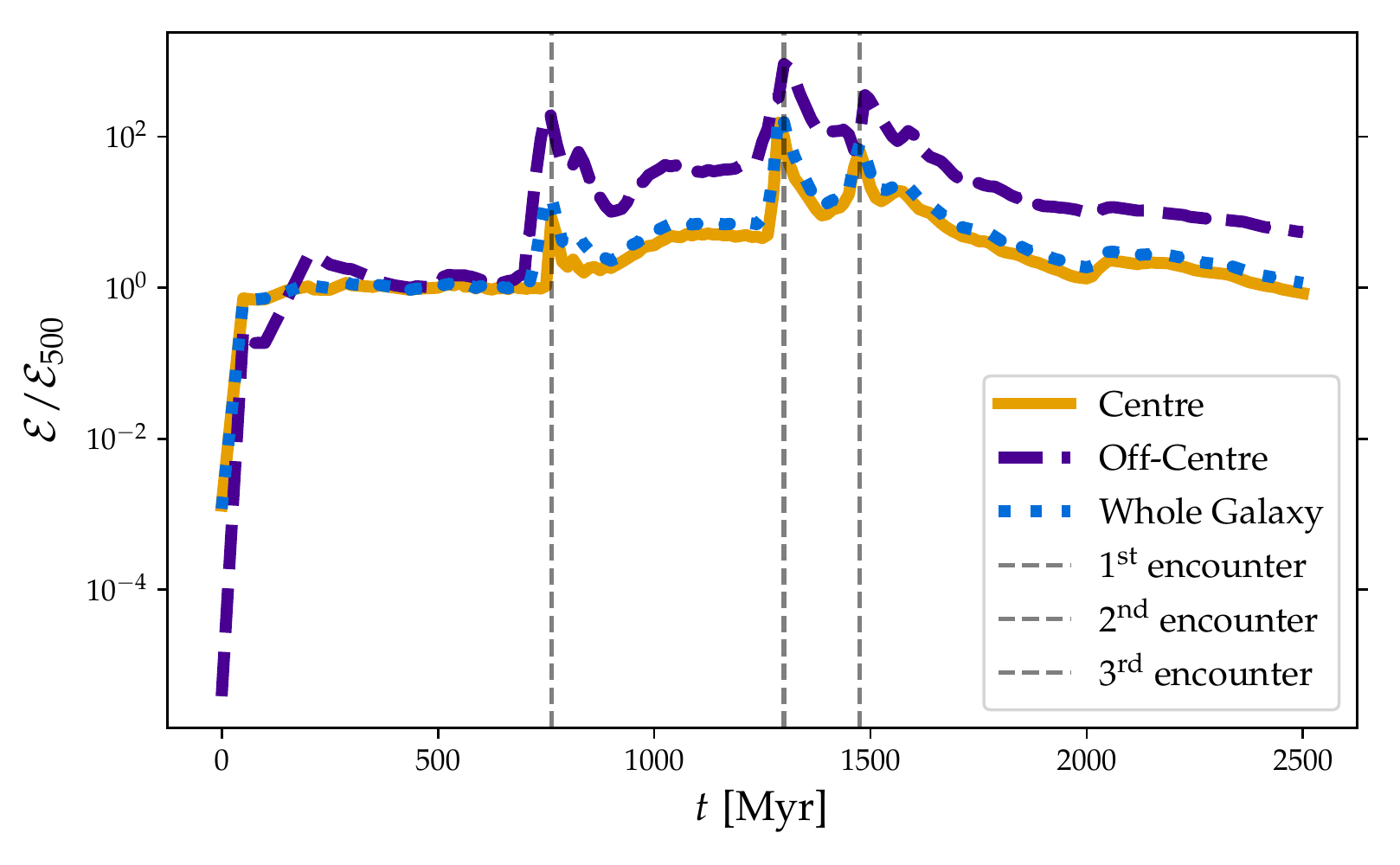}
  \includegraphics[scale=.4]{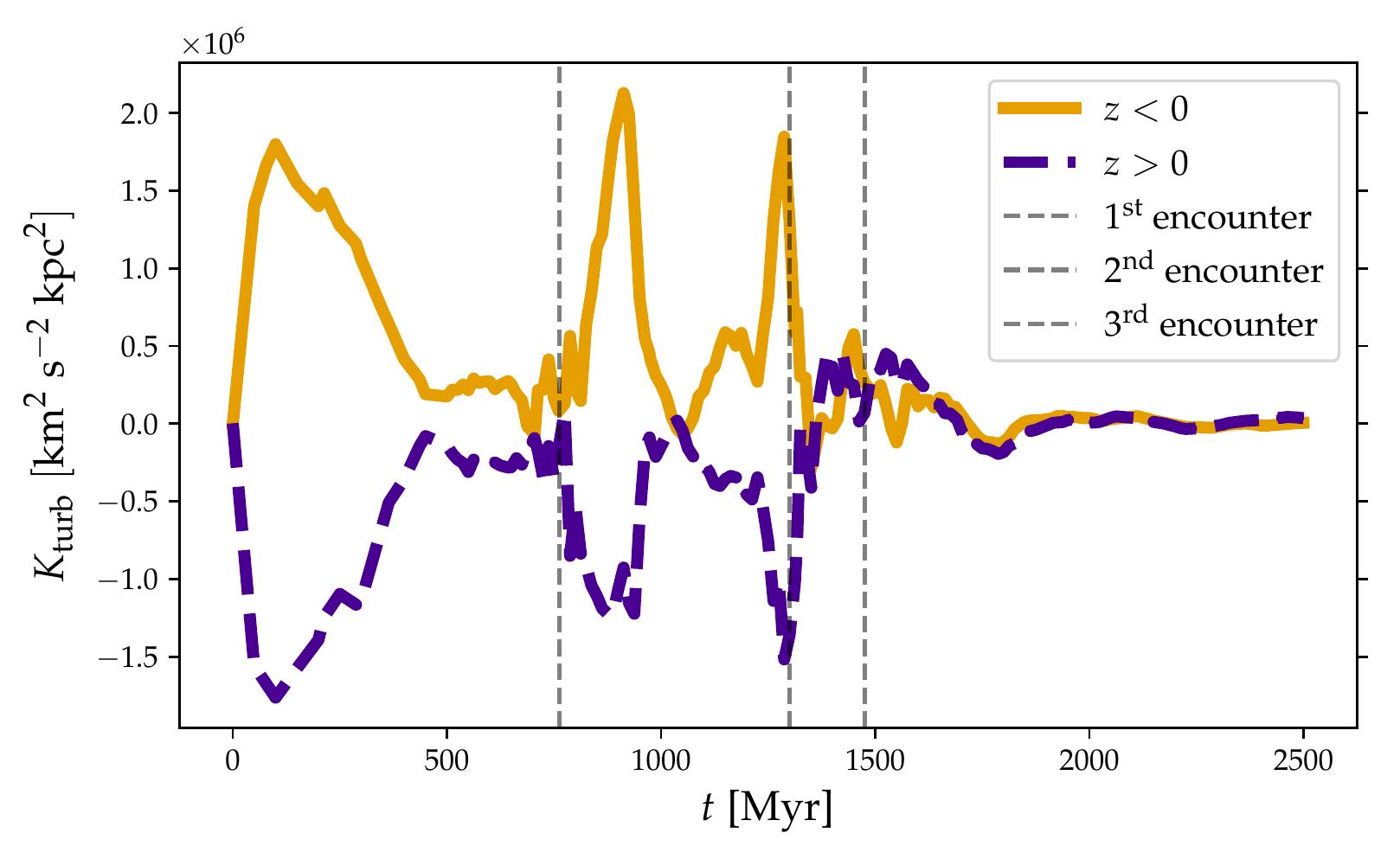}\\
  \caption{\textbf{Left}: Timeseries of the electromotive force $\mathcal{E}$ normalised by $\mathcal{E}_{500}$ in three regions of the primary galaxy. \textbf{Right}: Time evolution of the turbulent kinetic helicity above ($z>0$) and below ($z<0$) the primary galaxy's midplane. We highlight times of three encounters between the two galaxies (see Fig.~\ref{fig:MagneticFieldStrength}).} 
  \label{fig:emf+helicity}
\end{figure}

\begin{figure}[tb]
  \centerline{\vbox to 6pc{\hbox to 10pc{}}}
  \includegraphics[scale=.36]{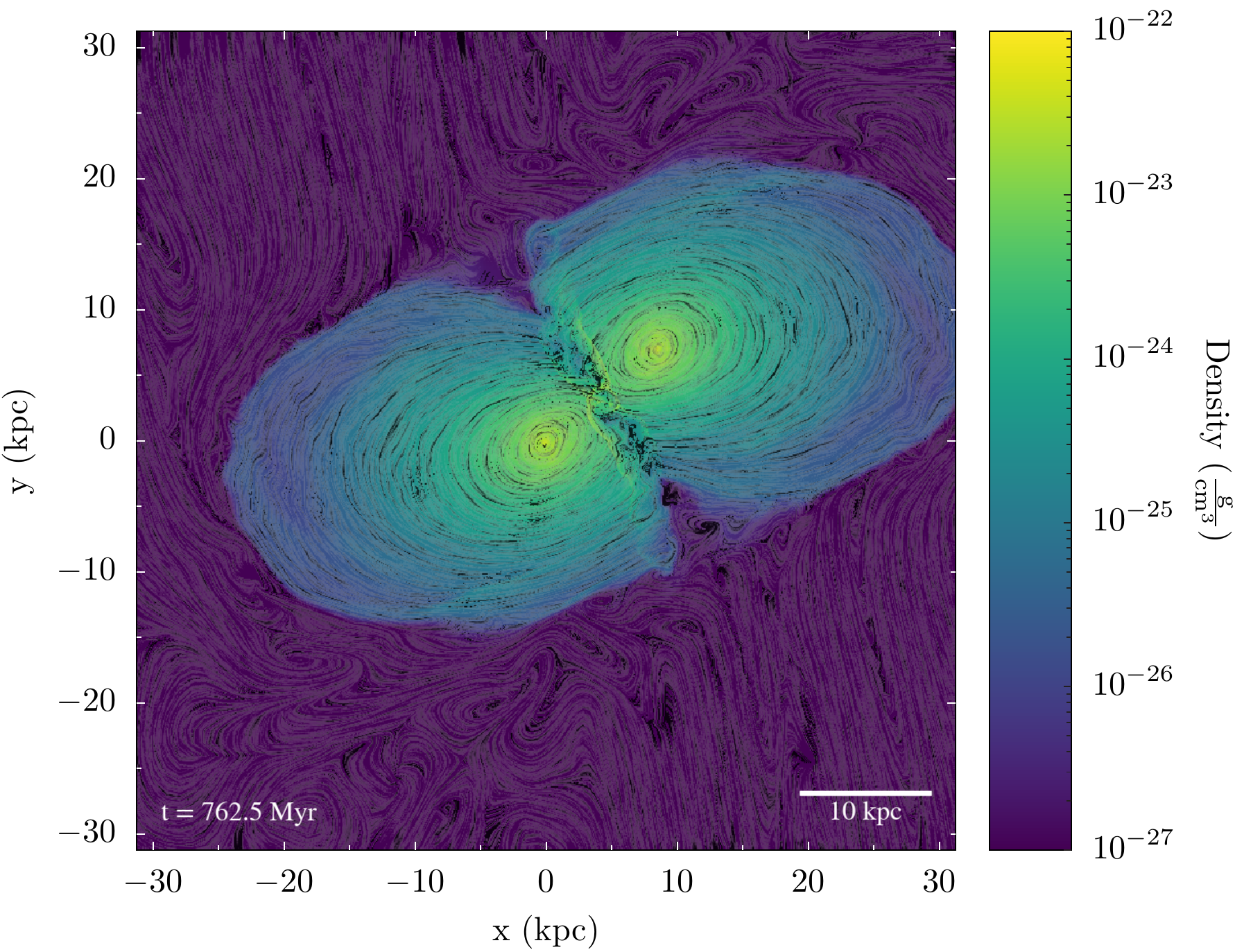}
  \includegraphics[scale=.36]{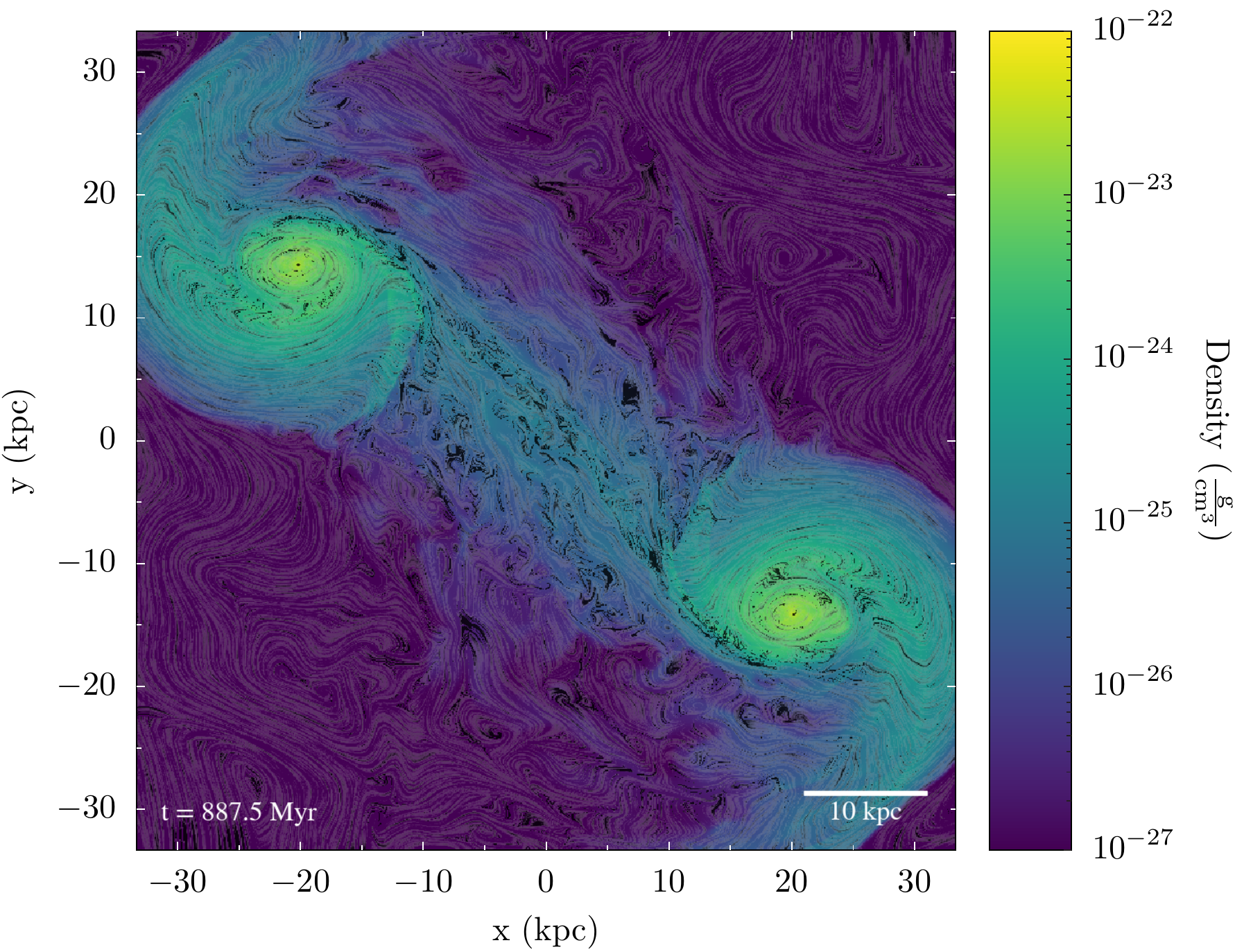}
  \caption{Two density slices along the midplane of the interacting galaxy pair overlaid with a representation of magnetic field lines provided by line integral convolution. The image on the left displays the first encounter and the image on the right the retracting galaxies between the first and second encounter.}
  \label{fig:slices}
\end{figure}

The magnetic field strength displayed in Fig.~\ref{fig:MagneticFieldStrength} experiences only a moderate rise at the first encounter where only the outer parts of the galaxies interact. Considerable amplification by a factor of $\sim 2$ at second and third encounters is accompanied by a stronger penetration of both galaxies. In Fig.~\ref{fig:slices} we show gas density slices along the disks' midplanes: the first encounter of the galaxies at $762.5\;$Myr and a snapshot at $887.5\;$Myr while they are moving apart. During the first encounter a shock front forms where the outer disks interact. Using an implementation of the line integral convolution method in yt \citep{CabralLeedom1993, Turk+2011} we illustrate the transition from ordered to random fields. At the first encounter the magnetic field is still toroidal outside of the collision zone while it appears random along the shock front. Furthermore, when the galaxies retract we can clearly recognise the turbulent pattern of the magnetic field in the bridge region. 

In order to study possible dynamo activity we apply mean-field electrodynamics \citep[e.g.][]{Parker1970, Steenbeck+1966} to our data in order to investigate the possible presence of a mean field dynamo \citep{Parker1971} in interacting galaxies. In mean-field electrodynamics both the magnetic field and the velocity of the fluid can be decomposed into fluctuating and mean components, respectively:
\begin{align}
    \mathbf{B}_{\mathrm{tot}}&=\overline{\mathbf{B}}+\mathbf{B}_{\mathrm{turb}} & &\& & \mathbf{v}_{\mathrm{tot}}&=\overline{\mathbf{v}} + \mathbf{v}_{\mathrm{turb}}. \label{eq:mfe}
\end{align}
To compute this we follow a similar approach as \citet{Ntormousi+2020}. To separate velocities and magnetic fields in the fashion of Eq.~\ref{eq:mfe} a mass-averaged filter with a scale of $\Delta x = 300\;$pc ($5^3$ cells) is applied. Substitution of Eq.~\ref{eq:mfe} into the induction equation and subsequent averaging it yields the mean-field induction equation
\begin{equation}
    \frac{\partial\overline{\mathbf{B}}}{\partial t} = \boldsymbol{\nabla}\times(\overline{\mathbf{v}}\times\overline{\mathbf{B}} + \mathcal{E}),\label{eq:dynamo}
\end{equation}
where we assume isotropic turbulence and ignore the magnetic diffusivity. We identify the electromotive force: 
\begin{equation}
    \mathcal{E} = \overline{\mathbf{v}_{\mathrm{turb}}\times\mathbf{B}_{\mathrm{turb}}}.
\end{equation}
Following e.g. \citet{Ntormousi+2020} $\mathcal{E}$ can be expanded in a series which yields after truncation
\begin{equation}
    \overline{\mathbf{v}_{\mathrm{turb}}\times\mathbf{B}_{\mathrm{turb}}} = \alpha\overline{\mathbf{B}} - \eta_{\mathrm{T}}\boldsymbol{\nabla}\times\overline{\mathbf{B}}\label{eq:dynamo2},
\end{equation}
where the transport coefficients $\alpha$ and $\eta_{\mathrm{T}}$ describe the $\alpha$-effect and the turbulent magnetic diffusivity, respectively.

Eq.~\ref{eq:dynamo2} describes the evolution of the averaged magnetic field. Comparing total and filtered fields in Fig.~\ref{fig:MagneticFieldStrength} we note that although peaks at second and third encounters are less pronounced, they still indicate significant amplification. 

We show the evolution of the electromotive force in Fig.~\ref{fig:emf+helicity}. We have normalised $\mathcal{E}$ by its value attained at $t=500\;$Myr which is still during the initial approach phase of the galaxies. Therefore, it is easier to observe changes in $\mathcal{E}$ with respect to the state of the galaxies before the first encounter when $\mathcal{E}/\mathcal{E}_{500}\approx 1$. Afterwards, the electromotive force keeps growing continuously until the third encounter, followed by  a gradual decline until pre-interaction values are reached in the central region towards the end of the simulation. Analogous to the magnetic field amplification observed in Fig.~\ref{fig:MagneticFieldStrength} events of interaction are marked by peaks with $\mathcal{E}/\mathcal{E}_{500}>10^{2}$ in the off-centre region.

In general, the transport coefficients $\alpha$ and $\eta_{\mathrm{T}}$ are tensors and complicated to evaluate. However, if we assume isotropic turbulence and neglect effects of Lorentz forces, $\alpha$ is proportional to the turbulent kinetic helicity \citep[see e.g.][and references therein]{Ntormousi+2020}
\begin{equation}
    K_{\mathrm{turb}} = \int{\mathbf{v}_{\mathrm{turb}}\cdot(\boldsymbol{\nabla}\times\mathbf{v}_{\mathrm{turb}})\mathrm{d}V}.
\end{equation}
The time evolution of the turbulent kinetic helicity in Fig.~\ref{fig:emf+helicity} presents us with three remarkable events where $K_{\mathrm{turb}}$ reaches maximum values thus offering particularly favourable conditions for a mean-field dynamo: (1) the initial phase $t<500\;$Myr, (2) the timespan between the first and the second encounters, and (3) the third encounter. The strong initial growth and the subsequent decrease of $K_{\mathrm{turb}}$ during the first 500 Myr is the result of a relaxation process within the gas disk. This is because during initialisation the DM halo is not adjusted to the gravitational potential of the gas disk. The second case could be associated with strong helical turbulence inside the tidal arms and the tidal bridge connecting the retracting collision partners. It is only at the third encounter where the maximum turbulent kinetic helicity coincides with the maximum magnetic field strength.

In summary, we find that while turbulence is induced by galaxy encounters and there is indication of a mean-field dynamo that also amplifies the magnetic field during these encounters, at the same time this amplification is not persistent. Magnetic field strength, turbulent kinetic helicity and the electromotive force are dissipated after the third encounter, suggesting that the dynamo is not sustained. 

We summarise the discussion following the presentation given at the symposium. A question addressed the transport coefficients in Eq.~\ref{eq:dynamo2} and whether we have computed $\eta_{\mathrm{T}}$. This is planned in future work.

The authors acknowledge funding from DFG grant SCHM 2135/6-1. The work was supported by the North-German Supercomputing Alliance (HLRN). We would like to thank the reviewer for their helpful comments.


\begin{thebibliography}{}
\bibitem[Beck(2015)]{Beck2015} Beck, R.\ 2015, {\it A\&A Rev.}, 24, 4  
\bibitem[Beck~{\it et al.}(1996)]{Beck+1996} Beck, R., Brandenburg, A., Moss, D., {\it et al.}\ 1996, {\it\araa}, 34(1), 155-206 
\bibitem[Binney \& Tremaine(2008)]{BinneyTremaine2008} Binney, J, Tremaine, S \ 2008, Galactic Dynamics 
\bibitem[Brandenburg \& Subramanian(2005)]{Brandenburg+2005} Brandenburg, A., Subramanian, K.\ 2005, {\it\physrep}, 417, 1-209
\bibitem[Bryan~{\it et al.}(2014)]{Bryan+2014} Bryan, G.~L, Norman, M.~L, O'Shea, B.~W., {\it et al.}\ 2014, {\it\apjs}, 211(2), 19 
\bibitem[Cabral \& Leedom(1993)]{CabralLeedom1993} Cabral, B., Leedom, L.~C.\ 1993, {\it Proceedings of the 20th Annual Conference on Computer Graphics and Interactive Techniques (New York)} 
\bibitem[Drakos~{\it et al.}(2017)]{Drakos+2017} Drakos, N.~E., Taylor, J.~E., Benson, A.~J.\ 2017, {\it\mnras}, 468(2), 2345-2358 
\bibitem[Drzazga~{\it et al.}(2011)]{Drzazga+2011} Drzazga, R.~T., {Chy{\.z}y}, K.~T., Jurusik, W., {\it et al.}\ 2011, {\it\aap}, 533, A22 
\bibitem[Freeman(1970)]{Freeman1970} Freeman, K.~C.\ 1970, {\it\apj}, 160, 811 
\bibitem[Grete~{\it et al.}(2019)]{Grete+2019} Grete, P., Latif, M.~A., Schleicher, D.~R.~G., {\it et al.}\ 2019, {\it\mnras}, 487(4), 4525-4535 
\bibitem[Grete~{\it et al.}(2017)]{Grete+2017} Grete, P., Vlaykov, D.~G., Schmidt, W., {\it et al.}\ 2017, {\it Phys. Rev. E.}, 95, 033206 
\bibitem[Helmi(2020)]{Helmi2020} Helmi, A.\ 2020, {\it\araa}, 58, 205-256 
\bibitem[Hernquist(1990)]{Hernquist1990} Hernquist, L.\ 1990, {\it\apj}, 356, 359 
\bibitem[Kotarba~{\it et al.}(2010)]{Kotarba+2010} Kotarba, H., Karl, S.~J., Naab, T., {\it et al.}\ 2010, {\it\apj}, 716(2), 1438-1452 
\bibitem[Kotarba~{\it et al.}(2011)]{Kotarba+2011} Kotarba, H., Lesch, H., Dolag, K., {\it et al.}\ 2011, {\it\mnras}, 415(4), 3189-3218 
\bibitem[Navarro~{\it et al.}(1996)]{Navarro+1996} Navarro, J.~F., Frenk, C.~S., White, S.~D.~M.\ 1996, {\it\apj}, 462, 563 
\bibitem[Ntormousi~{\it et al.}(2020)]{Ntormousi+2020} Ntormousi, E., Tassis, K., Del Sordo, F., {\it et al.}\ 2020, {\it\aap}, 641, A165 
\bibitem[Parker(1970)]{Parker1970} Parker, E.~N.\ 1970, {\it\apj}, 160, 383 
\bibitem[Parker(1971)]{Parker1971} Parker, E.~N.\ 1971, {\it\apj}, 163, 255 
\bibitem[Patton~{\it et al.}(2020)]{Patton+2020} Patton, D.~R., Wilson, K.~D., Metrow, C.~J., {\it et al.}\ 2020, {\it\mnras}, 494(4), 4969-4985 
\bibitem[Pillepich~{\it et al.}(2018)]{Pillepich+2018} Pillepich, A., Springel, V., Nelson, D., {\it et al.}\ 2018, {\it\mnras}, 473(3), 4077-4106 
\bibitem[Renaud~{\it et al.}(2019)]{Renaud+2019} Renaud, F., Bournard, F., Agertz, O., {\it et al.}\ 2019, {\it\aap}, 625, A65 
\bibitem[Renaud~{\it et al.}(2015)]{Renaud+2015} Renaud, F., Bournard, F., Duc, P.-A.\ 2015, {\it\mnras}, 446(2), 2038-2054 
\bibitem[Rodenbeck~{\it et al.}(2016)]{Rodenbeck+2016} Rodenbeck, K., Schleicher, D.~R.~G.\ 2016, {\it\aap}, 593, A89 
\bibitem[Schober~{\it et al.}(2013)]{Schober+2013} Schober, J., Schleicher, D.~R.~G., Klessen, R.~S.\ 2013, {\it\aap}, 560, A87 
\bibitem[Springel~{\it et al.}(2005)]{Springel+2005} Springel, V., Di Matteo, T., Hernquist, L.\ 2005, {\it\mnras}, 361 (3), 776-794 
\bibitem[Steenbeck~{\it et al.}(1966)]{Steenbeck+1966} Steenbeck, M., Krause, F., R\"{a}dler, K.-H.\ 1966, {\it Zeitschrift f\"{u}r Naturforschung A}, 21(4), 369-376 
\bibitem[Steinwandel~{\it et al.}(2019)]{Steinwandel+2019} Steinwandel, U.~P., Beck, M.~C., Arth, A., {\it et al.}\ 2019, {\it\mnras}, 438(1), 1008-1028 
\bibitem[Subramanian(2016)]{Subramanian2016} Subramanian, K.\ 2016, {\it Reports on Progress in Physics}, 79(7), 076901 
\bibitem[Turk~{\it et al.}(2011)]{Turk+2011} Turk, M.~J., Smith, B.~D., Oishi, J.~S., {\it et al.}\ 2011, {\it\apjs}, 192(1), 9 
\bibitem[Wang~{\it et al.}(2010)]{Wang+2010} Wang, H.-H., Klessen, R.~S., Dullemond, C.~P., {\it et al.}\ 2010, {\it\mnras}, 407(2), 705-720 



\end{thebibliography}
\end{document}